# Understanding Game Theory via Wireless Power Control

Giacomo Bacci, Luca Sanguinetti, and Marco Luise

In this lecture note, we introduce the basic concepts of game theory (GT), a branch of mathematics traditionally studied and applied in the areas of economics, political science, and biology, which has emerged in the last fifteen years as an effective framework for communications, networking, and signal processing (SP). The real catalyzer has been the blooming of all issues related to distributed networks, in which the nodes can be modeled as players in a game competing for system resources. Some relevant notions of GT are introduced by elaborating on a simple application in the context of wireless communications, notably the *power control in an interference channel (IC)* with two transmitters and two receivers.

## RELEVANCE

Recently, the mathematical tools of GT [1] have attracted a significant interest by the wireless communications and SP engineering communities [2, Part II], due to the need for designing autonomous, distributed, and flexible systems, in which the available resources are allocated through low-complexity and scalable procedures. Games are appealing, owing to some characteristics that are not common in classical optimization: as an example, GT can handle interactive situations in which each player can only have a partial control over the optimization variables, while using its own performance metric. It is true that commonalities can be found with other disciplines, such as multi-objective optimization [3], convex optimization [4], and learning theory [5], but GT possesses many distinguishing features that make it essential for the standard current toolbox of communication as well as SP engineers.

G. Bacci was with the University of Pisa, Pisa, Italy, and is now with MBI srl, 56121 Pisa, Italy (email: gbacci@mbigroup.it).

L. Sanguinetti and M. Luise are with the Dipartimento di Ingegneria dell'Informazione, University of Pisa, 56122 Pisa, Italy (email: luca.sanguinetti@unipi.it, marco.luise@unipi.it).

The research leading to these results has received funding from the European Union's FP7 under REA Grant agreements no. PIOF-GA-2011-302520 GRAND-CRU and PIEF-GA-2012-330731 Dense4Green, and is also supported by the European Union's FP7 NEWCOM# (Grant agreement no. 318306).

## Prerequisites

The readers require basic knowledge in linear algebra and wireless communication theory.

## What is a game?

To take advantage of GT and its associated theoretical tools, the first step is to model the problem at hand as a *game*. In doing so, three ingredients must be identified:

- the *players*, that represent the main actors in the problem, having conflicting interests and affecting the performance of everyone else in the game;

- a set of *strategies* available to each player, that determines what each player can do;

- a *utility* function for each player, that measures its degree of satisfaction as a function of the combination of all player's choices.

This description may encompass a large number of situations: to mention a few examples, players in a game can be base stations (BSs) allocating the resources in a cellular network to increase the system throughput, or watermarking devices choosing algorithms to face potential attackers.

The objective of the modeling effort is to describe the game using its *strategic-form representation*: a triplet $\langle \mathcal{K}, \{\mathcal{S}_k\}_{k \in \mathcal{K}}, \{u_k\}_{k \in \mathcal{K}} \rangle$, where: $\mathcal{K} = \{1, \ldots, K\}$ is the set of *players*, where $K$ is the number of players; $\mathcal{S}_k$ is the set of *strategies* for each player $k$; and $u_k(\mathbf{s})$ is the *utility function* (also known as *reward* or *payoff*) associated to player $k$ for a combination of choices $\mathbf{s} = [s_1, \ldots, s_K] = [s_k, \mathbf{s}_{\setminus k}]$, where $\mathbf{s}_{\setminus k} = [s_1, \ldots, s_{k-1}, s_{k+1}, \ldots, s_K]$ denotes the strategies taken by all other players except player $k$ (the *opponents*).

In general, the game outcome $u_k(\mathbf{s})$ for player $k$ depends on all players' choices through $\mathbf{s}$, that stems out from the interaction of the players with possibly conflicting interests. This brings forth a couple of distinguishing features of GT:

- each player $k$ can have a different performance metric; this feature is captured by a per-player specific function $u_k(\mathbf{s})$, that accounts for the player's nature;

- each player $k$ has partial control ($s_k \in \mathcal{S}_k$ only) over the optimization variables.

The first property is strictly tied with *multi-objective optimization* [3], although a clear difference exists in the scope of the optimization variables, as in multi-objective optimization we have full control over the variables. The second property is tightly related to the framework of *distributed optimization* [4], with which it shares many intersections, although there are specific differences: one of the most important is that, while in distributed optimization the agents follow some common given rules, in GT the players act as independent decision-makers.

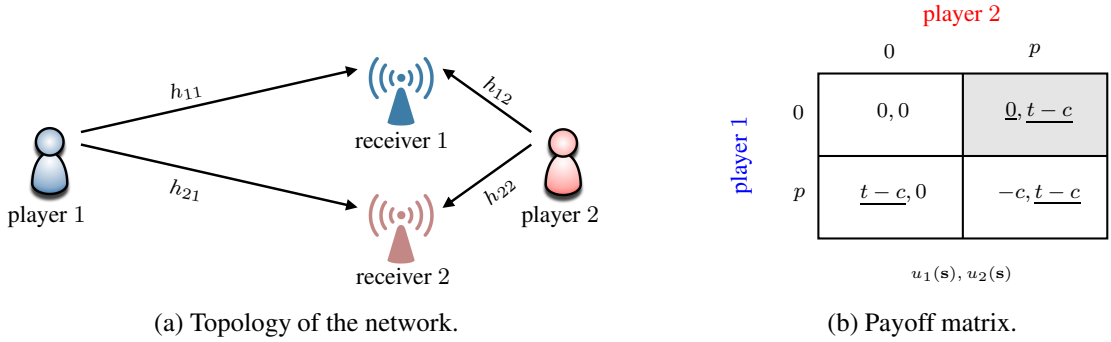

(a) Topology of the network.

(b) Payoff matrix.

Fig. 1: Representation of the NFE game.

## The near-far effect (NFE) game

To picture out the meaning of the strategic-form representation, let us consider an example taken from a very common wireless communications scenario: the IC, represented in Fig. 1a, in which the two transmitters interfere with each other in the attempt to reach their own receiver. This simple scheme encompasses many scenarios: it can be used to model *i*) a multicellular system, with red and blue nodes belonging to two different cells; *ii*) a heterogeneous network, where the red and blue nodes belong to a macro cell and a small cell, resp.; *iii*) a cognitive radio system, where the red and blue nodes are primary users (PUs) and secondary users (SUs), resp.; and *iv*) a device-to-device system, where the receivers are also network nodes.

Using GT, we can *i*) model the problem at hand in a suitable manner, and *ii*) provide the theoretical tools to solve it. In this case, *solving* means devising the optimal transmission strategy to be selected by the two wireless terminals of Fig. 1a. In particular, we assume that the two nodes are allowed either to transmit at a certain power level $p$, or to stay idle. This situation can be modeled as a game, with $K = 2$ players, and strategy sets $\mathcal{S}_k = \{0, p\}$ for $k \in \mathcal{K} = \{1, 2\}$. For simplicity, we also assume that the two terminals choose their strategies *simultaneously* (i.e., without being informed of the other's choice) *once and for all* (i.e., they cannot make any change after observing the outcome of the game) – in GT parlance, we consider a *static* game. Finally, since players and strategy sets are both countable, the game is *finite*.

As depicted in Fig. 1a, player 1 (the far terminal) is located much farther away from both receivers than player 2 (the near terminal). To describe this situation in a mathematical fashion, we introduce the *power gains* $h_{jk} \in \mathbb{R}^+$ experienced by terminal $k$'s signal when propagating to receiver $j$. For simplicity, let us assume $h_{jk} = h_k$ for $j = 1, 2$ with $h_1/h_2 \ll 1$ (we will better quantify this ratio later on), thereby giving rise to the so-called *near-far problem*.

We need now to define a utility function, and to do so we consider that each terminal achieves a degree of satisfaction that depends both on the success of its transmission and on the energy spent to transmit at power $s_k$. Mathematically, this translates here into a (dimensionless) utility $u_k(\mathbf{s}) = t_k(\mathbf{s}) - c_k(\mathbf{s})$, where $t_k(\mathbf{s})$ accounts for the outcome of the transmission, and $c_k(\mathbf{s})$ measures the cost associated to using $s_k$. We assume that the cost scales linearly with the transmit power, and that it is independent of the other terminal's strategy: $c_k(\mathbf{s}) = cs_k/p$. Measuring $t_k(\mathbf{s})$ is more complicated, as it has to capture the interaction between the players as a function of the selected strategies $s_1$ and $s_2$. In practice, successful reception of a signal in a multiple-access scenario (such as the one considered here), be it in the time, frequency, space, or code domain, depends on the signal-to-interference-plus-noise ratio (SINR) $\gamma_k$, which measures the ratio of the useful received signal power to the amount of undesired power collected at the receiver. Under the assumption of additive white Gaussian noise (AWGN) with power $\sigma^2$, we get

$$\gamma_k(\mathbf{s}) = \frac{\Gamma h_k s_k}{\sigma^2 + h_{\backslash k} s_{\backslash k}} = \mu_k(s_{\backslash k}) s_k \geq 0, \tag{1}$$

where $\backslash k = 2$ if $k = 1$, $\backslash k = 1$ if $k = 2$, and $\Gamma \geq 1$ is the processing gain, that depends on the multiple access technology and on the receiver processing. For the time being, let us assume the transmission to be successful if and only if $\gamma_k \geq \gamma_{\mathrm{req}}$, where the minimum SINR $\gamma_{\mathrm{req}}$ depends on some system parameters. So, when $\gamma_k < \gamma_{\mathrm{req}}$, the transmitted message cannot be decoded at receiver $k$, and $t_k(\mathbf{s}) = 0$. On the contrary, when $\gamma_k \geq \gamma_{\mathrm{req}}$, receiver $k$ can correctly receive the information associated to user $k$'s signal, and $t_k(\mathbf{s}) = t$, where $t$ is a dimensionless parameter that accounts for the throughput achieved at destination. To properly capture the cost-benefit analysis that regulates any practical wireless system, it makes sense to assume $t \gg c$.

A profitable way to investigate finite static games in their strategic form, such as our NFE game, is through the so-called *payoff matrix* (Fig. 1b) in which player 1's strategies are identified by the rows, player 2's strategies by the columns, and the entries of the matrix (the pair of numbers in the box) represent the utilities $(u_1(\mathbf{s}), u_2(\mathbf{s}))$ achieved by the players. Under the assumptions that $p = \frac{\sigma^2}{h_1 \Gamma} \gamma_{\mathrm{req}}$ and $\frac{h_1}{h_2} < \frac{1}{1 + \gamma_{\mathrm{req}}/\Gamma}$ (see [6] for further details), it is easy to fill out each box of the matrix based on the hypotheses listed above.

Once the game is in its strategic form, we have to *solve* it, i.e., to predict its outcome. In the NFE game, we assume that both players: *i)* are rational; *ii)* control their own strategies only; and *iii)* know each other's payoff. The first assumption means that each player is a utility-maximizer decision maker. The second hypothesis casts this problem as a *noncooperative* game, in which the players compete to unilaterally maximize $u_k(\mathbf{s})$. Finally, the third hypothesis involves the concept of *complete information* that each player has about the game. By inspecting the payoff matrix in Fig. 1b, it is apparent that player 2's best strategy is represented by $s_2^\star = p$ whatever $s_1$ is, since $t - c > 0$ under the assumption $t \gg c$. For this reason, the strategy $s_2 = 0$ is said to be *strictly dominated* by $s_2 = p$, as $u_2([s_1, 0]) < u_2([s_1, p]) \; \forall s_1 \in \mathcal{S}_1$. This is known to player 1 as well, that rationally chooses to play $s_1^\star = 0$. As a conclusion, the predictable outcome of the NFE game is $\mathbf{s}^\star = [s_1^\star, s_2^\star] = [0, p]$, as highlighted by the shaded box in Fig. 1b. In GT parlance, this game has been solved by applying the iterated elimination of dominated strategies, or *iterated dominance* for short [1, Ch. 1].

Fig. 2: Payoff matrix of the IC game.

## THE IC GAME

Let us now slightly modify the scenario represented in Fig. 1a. Assume for example to move player 1 closer to its receiver, such that the distance between player 1 and both receivers becomes the same as the distance between player 2 and both receivers. For simplicity, let us also suppose $h_{jk} = h$ for $j, k = 1, 2$. By using $p = \frac{\sigma^2}{h\Gamma} \gamma_{\text{req}}$, following the same considerations taken for the NFE game, it is easy to obtain the payoff matrix reported in Fig. 2. As an exercise, we can verify that no strictly dominated strategies exist, and thus we cannot apply the iterated dominance procedure used to solve the NFE game.

To get out of this *impasse*, we introduce the concept of *best response (BR)* $b_k(\mathbf{s}_{\setminus k})$, which is mathematically defined as:

$$b_k(\mathbf{s}_{\setminus k}) = \arg \max_{s_k \in \mathcal{S}_k} u_k \left( [s_k, \mathbf{s}_{\setminus k}] \right), \tag{2}$$

i.e., the best that we can get out of the game once we know the opponents' moves $\mathbf{s}_{\setminus k}$. Since player 1 chooses rows, we can compute its BR by examining the columns that can be possibly selected by player 2. When $s_2 = 0$, $b_1(s_2 = 0) = p$. Conversely, when $s_2 = p$, $b_1(s_2 = p) = 0$. The same can be obtained for player 2, and we end up with the players' BRs, underlining the relevant payoffs in Fig. 2. We find two boxes (shaded background) containing the BRs of both players, representing two *stable* states, where "stable" here means that such states are attained by some multiple agents with conflicting interests that compete through *self-optimization*, and eventually reach a point where none of them has any incentive to *unilaterally* deviate from.

A point that possesses such properties is termed a *Nash equilibrium (NE)* of the game, defined as a strategy profile $\mathbf{s}^\star = [s_k^\star, \mathbf{s}_{\backslash k}^\star]$ such that, for all $k \in \mathcal{K}$,

$$u_k\left([s_k^\star, \mathbf{s}_{\backslash k}^\star]\right) \geq u_k\left([s_k, \mathbf{s}_{\backslash k}^\star]\right) \qquad \forall s_k \in \mathcal{S}_k, \tag{3}$$

or, equivalently, $s_k^\star \in b_k(\mathbf{s}_{\backslash k}^\star)$. As an exercise, check that $\mathbf{s}^\star = [0, p]$ is the unique NE of the NFE game.

The notion of NE encompasses many interpretations of GT, not discussed here for brevity, that the interested readers can find in many textbooks (e.g., [7, Ch. 1]). Modeling the players as self-optimizing decision-makers finds a suitable application especially in the context of SP, in which the devices can be programmed to do so. Since each player has only a partial control of the game, the concept of NE is tightly coupled with the application of distributed algorithms and machine learning techniques [7, Part II].

For brevity, we will not discuss here theorems on *equilibrium existence* [1, Ch. 1], that establish the existence of the NE in particular classes of games, and on *equilibrium uniqueness* [2, Ch. 3]. When uniqueness cannot be ensured, like in the case of the IC game, we face the problem of *equilibrium selection*. One solution to this issue is the concept of *correlated equilibrium (CE)* [1, Ch. 2], a generalization of the NE, where an arbitrator helps the players to correlate their strategies, so as to favor a decision process in the interplay – e.g., letting them adopt $\mathbf{s}^\star = [p, 0]$.

### Introducing continuous powers

The solutions of NFE and IC games, in which at most one terminal can successfully transmit, directly stem out of choosing a binary strategy set $\mathcal{S}_k = \{0, p\}$ for both players. Let us see what happens if any power level in the continuous interval $[0, p]$ can be selected. This amounts to setting $\mathcal{S}_k = \{s_k \in \mathbb{R} : 0 \leq s_k \leq p\}$. Within this setting, the power control problem can be studied as a *continuous* game [7, Ch. 2]. In our attempt of getting closer to a realistic scenario, let us also modify the utilities to better model how real data networks work in practice. A good approximation for the effective throughput in a packet-oriented transmission is $t_k(\mathbf{s}) = t\,(1 - \exp\{-\gamma_k(\mathbf{s})\})^L$ [8], whose behavior is depicted (red line, left axis) in Fig. 3. In our expression, $L$ denotes the number of information bits per packet (here, $L = 20$), and $t$ is the communication rate (in b/s). To properly capture the tradeoff between obtaining a satisfactory throughput and saving transmit power – similarly to what considered for the NFE and IC games –, we will adopt a "green" approach, based on improving each player's energy efficiency [8]. This can be done by defining player $k$'s utility as the ratio between throughput and power expenditure, thus accounting for the number of

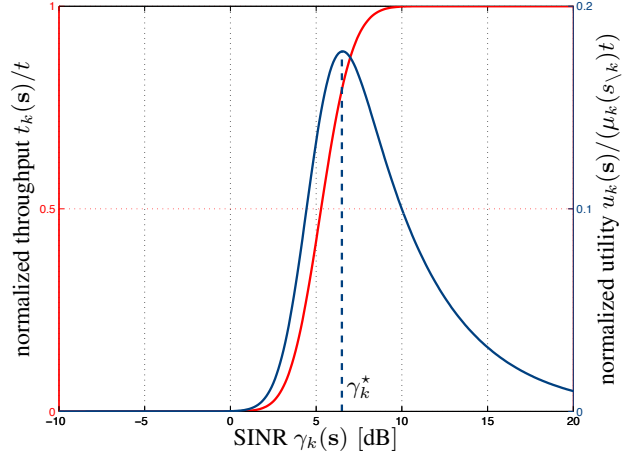

Fig. 3: Throughput (red) and utility (blue) as functions of the SINR (continuous NFE game).

bits correctly delivered per joule of energy consumed:

$$u_k(\mathbf{s}) = t_k(\mathbf{s})/s_k \qquad \text{[b/J]}, \tag{4}$$

whose normalized behavior is reported (blue line, right axis) in Fig. 3, with $\mu_k(s_{\setminus k})$ defined as in (1).

Using straightforward manipulation, player $k$'s BR (2) turns out to be $b_k(s_{\setminus k}) = \min\{p, \gamma_k^\star/\mu_k(s_{\setminus k})\}$, where $\gamma_k^\star$ is the "optimal" SINR such that $\partial t_k(\mathbf{s})/\partial \gamma_k(\mathbf{s})|_{\gamma_k(\mathbf{s})=\gamma_k^\star} = \frac{1}{\gamma_k^\star} t_k(\mathbf{s})|_{\gamma_k(\mathbf{s})=\gamma_k^\star}$ [8]. For instance, when $L = 20$, we get $\gamma_k^\star \approx 4.5 = 6.5\,\text{dB}$ for $k = 1, 2$ (see Fig. 3). Based on such BR, the continuous IC game presents a unique NE, represented by the fixed point $s_k^\star = b_k(s_{\setminus k}^\star)$ for $k = 1, 2$ [8].

How can the NE be "visualized"? Let us consider a particular realization of the network sketched in Fig. 1a, with the following parameters:[1] $h_{11} = 0.75$, $h_{21} = 0.25$, $h_{12} = 0.5$, $h_{22} = 1$; $\Gamma = 4$; $p/\sigma^2 = 5$; and $L = 20$. The solution of this game is given by the NE $\mathbf{s}^\star/\sigma^2 = [2.99, 1.97]$, yielding normalized utilities $\frac{\sigma^2}{t} u_1(\mathbf{s}^\star) = 0.269$ and $\frac{\sigma^2}{t} u_2(\mathbf{s}^\star) = 0.407$. We display in Fig. 4 the (normalized) utilities at the NE (green diamond) on the bidimensional normalized utility plane, given by all achievable utility pairs $(u_1(\mathbf{s}), u_2(\mathbf{s}))$ (shaded region), for any strategy profile $\mathbf{s} \in \mathcal{S}_1 \times \mathcal{S}_2$ (the utility plan can be found via a numerical search using [9]). Note that $s_1^\star > s_2^\star$, and $u_1(\mathbf{s}^\star) < u_2(\mathbf{s}^\star)$: this is due to the better channel conditions experienced by player 2 (both in the direct and the interference links), that make it achieve the optimal SINR $\gamma_2^\star$ with a lower power consumption than player 1. However, unlike the finite version of the NFE game, where $s_1^\star = 0$ (and thus $t_1(\mathbf{s}^\star) = 0$), now player 1 can successfully connect to its receiver, getting a throughput $t_1(\mathbf{s}^\star) \approx 0.8t$ (the same as $t_2(\mathbf{s}^\star)$), at the cost of a slightly higher power consumption ($s_1^\star/s_2^\star \approx 1.52$), and thus with a lower energy efficiency.

---

[1] The MATLAB code for all examples presented in this note is available for download in [9].

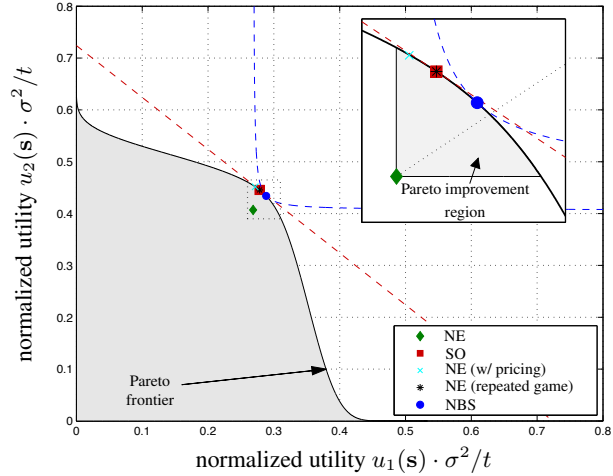

Fig. 4: Normalized utility plan (continuous-power NFE game).

### IS THE NE EFFICIENT?

A natural question that arises regards the actual efficiency, or, the performance, if you wish, of the NE. To address this question, we first need to agree upon our performance metric. In GT, a convenient way to assess how desirable a solution is involves the concept of *efficiency*, evaluated in terms of *Pareto optimality*. A profile $\overline{\mathbf{s}}$ is Pareto-optimal (PO) if there exists no other $\mathbf{s}$ such that *i)* $u_k(\mathbf{s}) \geq u_k(\overline{\mathbf{s}})$ for all $k \in \mathcal{K}$, and *ii)* $u_k(\mathbf{s}) > u_k(\overline{\mathbf{s}})$ for some $k \in \mathcal{K}$. In our continuous IC game, the performance achieved by the PO profile set is represented by the contour of the shaded area of Fig. 4, which is called the *Pareto frontier*. Clearly, if we increase $u_1(\mathbf{s})$ (i.e., if we move rightward along $x$), then $u_2(\mathbf{s})$ decreases, and the same happens if we increase $u_2(\mathbf{s})$ (by moving upward along $y$). Still, Pareto optimality does not qualify as our performance metric.

We have to further introduce the notion of social welfare (SW), that is often used as a convenient measure for the efficiency of a strategy vector [7, Ch. 2]. Formally, the *social-optimal (SO) profile* $\breve{\mathbf{s}}$ is the PO profile that maximizes the SW, defined as the weighted sum-utility $\sum_{k \in \mathcal{K}} w_k u_k(\mathbf{s})$, where the weights $\{w_k\}_{k \in \mathcal{K}}$, $\sum_{k \in \mathcal{K}} w_k = 1$, allow us to account for different classes of service: as an example, unequal weights can be useful to model PUs (higher $w_k$'s) and SUs (lower $w_k$'s) in a cognitive network. In our two-player game, we can identify $\breve{\mathbf{s}}$ as the tangent point between the Pareto frontier and a line with slope $-w_1/w_2$. As an example, if we consider $w_1 = w_2 = 0.5$ (i.e., if the two players have the same priority), $\check{s}/\sigma^2 = [2.20, 1.55]$, yielding $\frac{\sigma^2}{t}u_1(\check{s}) = 0.278$ and $\frac{\sigma^2}{t}u_2(\check{s}) = 0.446$ (Fig. 4).

We have zoomed a section of Fig. 4 (see the inset box) to show that the NE $s^\star$ is socially inefficient, since its performance is distinct from (more specifically, poorer than) that achieved by $\check{s}$. In general, $\check{s}$ *cannot* be achieved by distributed algorithms, rather, it is the result of a global optimization, and in our case it turns out to be unbalanced towards player 2 (the one with better channel conditions): this is reminiscent of the waterfilling policy [4], that allocates most resources to the users that can achieve higher throughputs. More importantly, the magnification in Fig. 4 shows that there are a multitude of profiles that provide utilities lying in the *Pareto improvement region*, in which $u_k(s) \geq u_k(s^\star)$ for all $k \in \mathcal{K}$ (shaded region in the inset). Consequently, the next question is: how can we improve the efficiency of the NE? In this lecture note, we will focus on three popular methods, namely: *i)* modifying the utility functions; *ii)* letting the players interact more than once; and *iii)* letting the players cooperate.

## Pricing the strategies

The simplest method to improve the efficiency of the NE while maintaining the game structure is by modifying the utility function. This can be done for instance by introducing some form of *externality*. This approach is in spirit close to mechanism design [1, Ch. 7]. For the power control games studied so far, one might think of charging the players for the powers they consume, by introducing a *pricing factor* $\alpha$ (in b/J · W$^{-1}$): $\tilde{u}_k(s) = u_k(s) - \alpha s_k = t_k(s)/s_k - \alpha s_k$ [10]. The rationale behind this approach is the following: if each transmitter is discouraged from being aggressive (due to power taxation), the multiple access interference (MAI) experienced by the other is reduced, and both SINRs at the equilibrium stay as close as possible to $\gamma_k^\star$ (provided that the AWGN power $\sigma^2$ is not dominant – a condition which always holds in multiple-access systems).

To evaluate the benefits of this method, we compute the NE $\tilde{s}^\star$ of the modified game, using the BR approach (see [9]). The cyan cross marker in Fig. 4 represents the performance $\left( \frac{\sigma^2}{t}u_1(\tilde{s}^\star), \frac{\sigma^2}{t}u_2(\tilde{s}^\star) \right)$ of the *original* utility function (4) (which is the one we are actually interested in), using $\alpha = 0.12t/\sigma^4$, that yields $\tilde{s}^\star/\sigma^2 = [2.17, 1.57]$. As can be seen, the performance of $\tilde{s}^\star$ is very close to the Pareto frontier, further favoring player 2's performance compared to the SW. We have thus improved the efficiency of the game solution, while maintaining the noncooperative nature of the interplay (with all its desirable properties). The main drawback is that this improvement can only be achieved after a proper tuning of $\alpha$ (that highly depends on system parameters). As an exercise, one could evaluate the performance of $\tilde{s}^\star$ as a function of $\alpha$ (see [9]).

## Repeating the game

The inefficiency of the NE is mainly due to the selfish behavior (in the sense of self-optimization) of the players. An effective method to induce *cooperation* – while maintaining the noncooperative nature of the interaction – is forcing the players to interact more than once. A typical example of this approach is a *repeated game*, in which a static game is repeated $N$ times. For instance, assume that the two transmitters of Fig. 1a interact a number $N$ of times, each time selecting their optimal transmit powers $s_k(n)$, where $n$ is the time index [11]. When introducing the notion of time, each strategy set $\mathcal{S}_k$ becomes a *complete plan of actions*, that depends on the unfolding of the game through time.

Similarly, the utility functions must account for *i)* the partial utilities $u_k(\mathbf{s}(n)) = t_k(\mathbf{s}(n))/s_k(n)$ received at each stage $n$ of the game, with $\mathbf{s}(n)$ denoting the profile selected by the players at time $n$; and *ii)* how much past utilities should be weighted (i.e., decay) compared to present ones. A simple example is the exponential decay, where the utility at time $n$ is weighted by the factor $\delta^n$, $0 \leq \delta \leq 1$, and the total utility after $N$ repetitions of the game is $u_k^N(\mathbf{s}) = \sum_{n=0}^{N} \delta^n u_k(\mathbf{s}(n))$. By letting $N \to \infty$, we further define a normalized utility $u_k^\delta(\mathbf{s}) = (1-\delta) \sum_{n=0}^{+\infty} \delta^n u_k(\mathbf{s}(n))$ [11]. The parameter $\delta$ is the so-called *discount factor*, and its meaning is borrowed from micro-economics: a payoff received at the present time $n$ is larger by a factor $1/\delta$ than the payoff of the next stage, and smaller by a factor $\delta$ than that of the previous one. This means that, if players are patient (in the SP and communications context, delay-tolerant), $\delta$ is typically close to 1. Conversely, if players are impatient (i.e., delay-sensitive), $\delta$ is typically close to 0.

Extending the concept of NE to repeated games [1, Ch. 5], we can show that the optimal strategy $\mathbf{s}^\delta = \{\mathbf{s}^\delta(n)\}_{n=0}^{+\infty}$ is for both players to select $s_k^\delta(n) = \breve{s}_k$ if $\left(s_k^\delta(n-1) = \breve{s}_k \text{ and } s_{\backslash k}^\delta(n-1) = \breve{s}_{\backslash k}\right)$, and $s_k^\delta(n) = s_k^\star$ otherwise, with $s_k^\delta(0) = \breve{s}_k$, and $\breve{\mathbf{s}}$ and $\mathbf{s}^\star$ being the SO and NE points, respectively, provided that $\delta \geq \underline{\delta}$ (i.e., if they are delay-tolerant enough), where $\underline{\delta}$ is a function of the network parameters [11]. In other words, in the repeated IC game, cooperation is enforced by letting the players interact an indefinite number of times: this is successful due to threatening future punishments for the player(s) who defect.

The effectiveness of this approach is apparent in Fig. 4, where the performance of $\mathbf{s}^\delta$, represented by the black asterisk, coincides with the SW, under the assumption $\delta \geq \underline{\delta}$. As a drawback, players must have knowledge of $\breve{\mathbf{s}}$: this only occurs if each player $k$ knows all channel gains $\{h_{jk}\}_{j,k \in \mathcal{K}}$, which might not be viable for all scenarios (e.g., in a cognitive network). Repeated games are a subclass of *dynamic games*, that are often used in SP problems to account for time evolution (see [7, Ch. 3] for more details).

## Introducing cooperation among the players

In the techniques considered so far, we have focused on improving the efficiency of the solution, without considering any *fairness* issue. In the example above, the SW is obtained by favoring player 2 to the detriment of player 1's performance, as is apparent in the inset of Fig. 4, where the SW is far away from the projection of the NE over the Pareto frontier – obtained intersecting it with the "fair" line with slope 1 and passing through $(u_1(\mathbf{s}^\star), u_2(\mathbf{s}^\star))$. We can balance efficiency and fairness by explicitly introducing *cooperation* among the players, assuming some explicit exchange of information. The fundamental difference of a cooperative approach is that, while in the games assumed so far cooperation can only be induced as the result of matching it with self-optimization (i.e., unilateral deviations are not beneficial anyway), now the players are *willing* to cooperate, as they know that they can mutually benefit from reaching an agreement. In GT parlance, this is called a *bargaining problem* [2, Ch. 7], whose analytical tools are tightly related to SP techniques, such as consensus algorithms [12].

Consider again the continuous IC game, and assume that the players can collaborate to select a satisfactory profile $\dot{\mathbf{s}} \in \mathcal{S}_1 \times \mathcal{S}_2$. In case they fail to reach an agreement, each player $k$ gets $u_k(\mathbf{s}^\star)$, where $\mathbf{s}^\star$ is the NE of the noncooperative game studied before. On the contrary, the players now strive to attain the *Nash bargaining solution (NBS)*, i.e., the (unique) PO profile that satisfies $\dot{\mathbf{s}} = \arg \max_{\mathbf{s} \in \mathcal{S}} \prod_{k=1}^{2} (u_k(\mathbf{s}) - u_k(\mathbf{s}^\star))$, where the subset $\mathcal{S} \subseteq \mathcal{S}_1 \times \mathcal{S}_2$ is such that $u_k(\mathbf{s}) \geq u_k(\mathbf{s}^\star)$ for all $k \in \mathcal{K}$ and $\mathbf{s} \in \mathcal{S}$. Interestingly, the NBS has close analogies with proportional fair allocation mechanisms, as discussed in [2, Ch. 7]. As is apparent, the NBS tries to increase as much as possible the utilities of the players with respect to the NE in a fair manner.

The graphical interpretation of the NBS is shown in Fig. 4: the NBS $\dot{\mathbf{s}}$ corresponds to the profile such that $(u_1(\dot{\mathbf{s}}), u_2(\dot{\mathbf{s}}))$ is the point of tangency between the Pareto frontier and the hyperbola with vertex in $x = u_1(\mathbf{s}^\star)$ and $y = u_2(\mathbf{s}^\star)$. Hence, the point of tangency lies by definition in the Pareto improvement region, as illustrated in Fig. 4 using a blue dot. In our usual network configuration, $\dot{\mathbf{s}}/\sigma^2 = [2.26, 1.52]$, yielding $\frac{\sigma^2}{t} u_1(\dot{\mathbf{s}}) = 0.288$ and $\frac{\sigma^2}{t} u_2(\dot{\mathbf{s}}) = 0.434$. The performance of $\dot{\mathbf{s}}$ lies in between the SW and the maximum-fairness projection of the NE performance, thus trading off efficiency and fairness. The reason why the NBS is unbalanced towards player 2 lies again in its better channel condition, which makes it stronger in negotiation [2, Ch. 7].

For more than two players, we can also consider a more general cooperative framework, namely *coalitional GT* [13], that provides the theoretical tools to investigate situations in which subsets of players can bind agreements to work together, aiming at improving their joint utility. This approach is particularly useful in many areas of SP, such as spectrum sensing for cognitive systems (see [2, Ch. 13] for further details).

## CONCLUDING REMARKS

In this lecture note, we introduced the very basic notions of GT, using a power control problem for a wireless interference channel as the *leitmotiv*: by further detailing and adding features to this "toy example," we presented, among the others, the concepts of players, strategies, utilities, NE, Pareto and social optimality. The interested readers that want to deepen their knowledge of GT are invited to target specific textbooks, such as general ones (e.g., [1]), and those specifically tailored to an SP audience (e.g., [2], [7]).

## REPRODUCIBLE RESEARCH

This lecture note has supplementary, downloadable material available in [9], provided by the authors. The material includes MATLAB code that can reproduce all the simulation results.